\documentclass{article}
\usepackage{conf,amsmath,epsfig,epsbox}

\title{ 
Quantum Gauged Neural Network: U(1) Gauge Theory}

\name{
Yukari Fujita ${}^*$\thanks{${}^*$ e-mail address: 
yukari@phys.kindai.ac.jp}    and
Tetsuo Matsui ${}^{\dagger}$\thanks{${}^\dagger$ 
e-mail address: matsui@phys.kindai.ac.jp}}
\address{Department of Physics, Kinki University, Higashi-Osaka, 
Japan 577-8502}

\begin{document}

%
\maketitle
\begin{abstract}
A quantum model of neural network is introduced and its phase structure
is examined.
The model is an extension of the classical Z(2) gauged neural network
of learning and recalling to a quantum model by replacing
the Z(2) variables, $S_i = \pm1$ of neurons and $J_{ij} =\pm1$ 
of synaptic connections, to the U(1) phase variables, 
$S_i = \exp(i\varphi_i)$ and $J_{ij} = \exp(i\theta_{ij}) $.
These U(1) variables describe the phase parts of 
the wave functions (local order
parameters) of neurons and synaptic connections. 
The model takes the form similar to
the U(1) Higgs lattice gauge theory, the continuum limit 
of which is the well known Ginzburg-Landau theory 
of superconductivity. Its
current may describe the flow of electric voltage along axons and 
chemical materials transfered 
via synaptic connections. The phase structure of the model at 
finite temperatures is examined by the mean-field theory, and
Coulomb, Higgs and confinement phases are obtained.
By comparing with the result of the $Z(2)$ model, the quantum effects 
is shown to weaken the ability of learning and recalling.
\end{abstract}
\section{Introduction}
\label{sec:intro}

To study rich activities of human brains, there are various
approaches. A typical one is neural networks. 
Various models of neural networks have been proposed.
The Hopfield model of associative memory\cite{hopfield} has offered
us a good explanation of the mechanism how we recall patterns.
On the other hand, the perceptron or its improvement, 
the back-propagation 
model\cite{perceptron}, may be a representative model of learning.

In Ref.\cite{matsui,kemuriyamaandmatsui}, yet another 
network model is proposed, which is an extension of the Hopfield
model to a model of learning by treating the strength $J_{ij}$ of
the synaptic connection between $i$-th and $j$-th neurons as 
an independent dynamical variable. Both the neuron variables
$S_i = \pm1$ and the new variables $J_{ij} = \pm1$ 
are treated on an equal footing. 
$J_{ij}$ is viewed as a ``connection" of gauge theory\cite{wilson}, 
and the energy $E(\{S_i\}, \{J_{ij}\})$ is postulated
to possess the local $Z(2)$ gauge symmetry. The gauge symmetry
assures us that the time evolutions of $S_i$ and $J_{ij}$ occur through local
(contact) interactions as they should be.

Among approaches other than neural networks, there is a  
quantum-theoretical approach to the brain activities.
Stuart, Takahashi and Umezawa\cite{umezawaandtakahashi}
proposed a microscopic quantum field theory by 
using operators expressing neurons and intermediating bosons.
They proposed that memory should be stored in low-energy modes
like Goldstone bosons. Jibu and Yasue\cite{jibuandyasue} argued 
that their quantum brain model may be regarded as a practical model 
of dipoles of ordered water and evanescent (massive) 
photons in the brain. 
 
Another quantum approach is advocated by 
\newline
Penrose. \cite{penrose} 
He insists on
the relevance of quantum theory like the problem of 
observations in quantum mechanics, coupling to quantum gravity, and so on.
It seems ambitious, but interesting and worth enough to scrutinize its 
validity. 
Hameroff and Penrose\cite{hameroffandpenrose} proposed a quantum theory 
 of consciousness.
They claim that objective reductions of wave functions of
microtubules, main building blocks of axons connecting
neurons, are relevant for our consciousness.
The central physical quantity in their theory is the so 
called decoherence time $\tau$, the average time interval between
successive reductions. $\tau$ corresponds to each ``moment"
of the stream of one's consciousness. There are several estimates 
of $\tau$\cite{tegmark,matsuiandsakakibara}, 
but they seem to be still controvercial each other. 
 
In this paper, we introduce a quantum version of the gauged neural
network of learning and recalling\cite{matsui,kemuriyamaandmatsui}.
This quantum neural network is regarded as an effective (phenomenological)
model  at macroscopic scales derived from 
the underlying microscopic quantum theory of brain.
The purpose of this neural network model 
is to explore the difference between classical and 
quantum neural networks and eventually to find the possible relevance
of quantum natures in the activities of human brains.
The structure of the paper is as follows;
In Sect.\ref{sec:model}, we introduce the quantum gauged model.
In Sect.\ref{sec:phase}, we study the phase structure of the model
at finite ``temperatures" $T$. 
In Sect.\ref{sec:discussion} we present conclusions and future
problems.

\section{Quantum Gauge Model}
\label{sec:model}

In this section, we first explain the relevance of gauge symmetry.
Next we discuss the possible ways to include quantum effects.
Then we propose an explicit model and the rule of time evolution.  

\subsection{Gauge Symmetry}

In the Hopfield model, the state of $i$-th neuron (active or inactive) 
is described
by the $Z(2)$ variable $S_i ( = \pm 1)$, and the state of the synaptic 
connection between $i$-th and $j$-th neurons is expressed
by its strength $J_{ij}$, which is a preassigned constant.  
The signal at the $j$-th site at time $t$, 
$S_j(t)$, propagetes to the $i$-th site through the axon and 
 the synaptic connection in the form $V_{ij}S_j$ to affect
 the state $S_i$ in the next time step $t + \Delta t$;
\begin{eqnarray}
S_i(t + \Delta t) = sgn\left[\sum_{j}J_{ij}S_j(t)\right].
\label{timeevolution}
\end{eqnarray}
The time evolution (\ref{timeevolution}) is known to decrease 
(not increase) the  ``energy", 
\begin{eqnarray}
E &=& -\frac{1}{2} \sum_{i,j} S_{i} J_{ij} S_j.
\label{hopfieldenergy}
\end{eqnarray}

In order to study processes of learning certain patterns of $S_{i}$,
it is necessary to allow for the time variation of $J_{ij}$.
There are various proposals how to treat the dynamics of $J_{ij}$.
The idea in Ref.\cite{matsui,kemuriyamaandmatsui} is to regard
$J_{ij}$ as a connection variable $U_{xy}$  of gauge theory\cite{gauge}. 
This is quite natural
because the connection describes the way how a pair of two points
 are connected, i.e., how two internal coordinates are related.  
In fact, $U_{xy}$ transports a quantity $\varphi_y$ (e.g., a vector) 
at a point $y$
to another point $x$ via the ``parallel-translate, the result
being $U_{xy}\varphi_y$. (See Fig.1.) 
One may view the signal $J_{ij}S_j$ in the Hopfield model 
as the result of parallel- 

\begin{center}
\postscriptbox{6cm}{5cm}{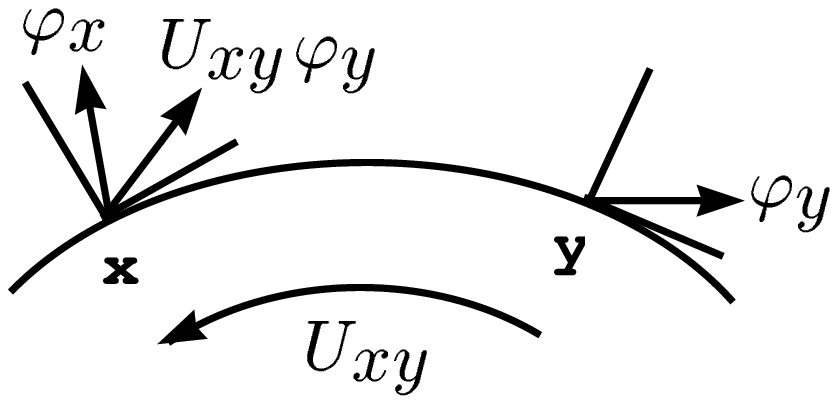}
\end{center}

\vspace{-2.cm}
Fig.1.
\ Function of gauge connection $U_{xy}$.
$U_{xy}$ parallel-translates a vector $\varphi_y$ at the point $y$ 
to another point $x$ giving rise to $U_{xy}\varphi_y$. To compare
$\varphi_y$ with a vector $\varphi_x$ at $x$, one should take the gauge
invariant scalar
product $(\varphi_x, U_{xy} \varphi_y) \equiv \varphi^\dagger_x U_{xy} 
\varphi_y$ instead of $(\varphi_x, \varphi_y) =\varphi^\dagger_x \varphi_y$.
\label{fig:gauge}

\newpage
\noindent
translating  $S_j$ to the $i$-th site by regarding
$J_{ij} \rightarrow U_{xy},$ $\ S_j \rightarrow \varphi_y$.
Since the connection is an independent variable in nature,
$J_{ij}$ is no more a constant but should be time-dependent. 
Both  $S_i(t)$ and $J_{ij}(t)$ should be treated
on an equal footing.
Once the system is regarded as a gauge system, it
should possess the gauge symmetry. That implies the
energy $E(\{S_i\}, \{J_{ij}\})$ is invariant under gauge transformations.

To be general, let us consider a gauge group $G$.
Then we prepare gauge variables $J_{ij} \in G$ (unitary representation
(unitary matrix) of a group element) for each pair $i,j$
and neuron variables $S_i \in G$ (fundamental representation (vector)) 
for each point $i$.
The gauge symmetry is local, that is the following gauge transformation 
can be performed at each point $i$ independently;
\begin{eqnarray}
S_i &\rightarrow& S_i^{'} \equiv V_i S_i, 
\nonumber\\
 J_{ij} &\rightarrow& J_{ij}^{'} \equiv V_{i} 
 J_{ij} V^\dagger_{j},\nonumber\\
 V_i & \in& G, \ 
 \end{eqnarray}
where $V_i$ is a unitary matrix ($V_i^\dagger = V_i^{-1}$).
The gauge invariance of $E$ is expressed as
\begin{eqnarray}
E(\{S'_i\}, \{J'_{ij}\}) &=& E(\{S_i\}, \{J_{ij}\}).
\end{eqnarray}
A simple example of the energy is
\begin{eqnarray}
E &=& -\frac{1}{4}\sum_{i,j}\left( S^\dagger_{i} J_{ij} S_j 
+ {\rm c.\ c}\right).
\end{eqnarray}
Note that the gauge invariance of $E$ holds at $j$, for example,  
since $V_{j}$ supplied by $S'_j$
cancels with $V^\dagger_j$ supplied by $J'_{ij}$ 
(note $V^\dagger_j V_j = 1$). 
For $G=Z(2)$, this $E$ reduces to
the Hopfield energy (\ref{hopfieldenergy}).

To consider generalization of the energy, the principle of 
gauge symmetry puts severe restrictions on $E$. 
Actually, the gauge principle implies that the time
evolutions of $S_i$, $J_{ij}$, which we shall discuss 
in Sect.\ref{ssec:timeevolution} in details,  
are controlled only by those 
signals  that have contacts with them.
For example, $dS_i/dt$ consists of terms, each of which has 
the index $i$ like  $V_{ij}S_j$. 
This assures us that the flows (current) of electric voltages and
chemical materials  change locally
through contact intereactions as it should be.

\subsection{Quantum Effects}
\label{ssec:quantumeffects}

Most of the proposed models of neural networks so far is classical
in the sense that these models employ real numbers as their
dynamical variables. 
Although there are many successful 
phenomenological models in the framework of classical physics
in various fields of physics, every physical system is necessarily 
``quantum" in its origin.
Neural networks are not an exception at all.

From the microscopic point of view,  main functions of our brains
should be the result of underlying microscopic systems, basic
constituents of which are electrons and various chemical materials.
The quantum brain theory of 
Stuart, Umezawa and Takahashi\cite{umezawaandtakahashi}
may be viewed as such a microscopic model.\cite{jibuandyasue}
As another approach, the recent quantum-theoretical study of 
consciousness by Hameroff and 
Penrose \cite{hameroffandpenrose} are also interesting since they focus 
on a microtubule and
start form its microscopic model itself. Actually, they consider
a two-dimensional system of electrons and its wave function.
The time dependence of wave function, particularly
its objective reductions,  is argued to be important for understanding
consciousness. In Ref.\cite{matsuiandsakakibara} 
a quantum-field-theoretical model of a microtubule is proposed, 
Hamiltonian of which is described by second-quantized fermionic 
electron operators. The model resembles familiar strongly-correlated 
electron systems like Hubbard model, Heisenberg model, t-J model, etc.

It is quite interesting to compare these quantum models and
existing classical neural-network models to identify the quantum effects.
However, to perform such a comparison explicitly, the present forms
of these quantum models are not appropriate; they involve quantum 
operators and have complicated structures.
Thus it is preferable
to obtain their effective models (at lower energies, i.e., at 
macroscopic scales) that take forms similar
to the classical neural networks. (See Fig.2.)

At this point, we recall the relation between
the BCS (Bardeen-Cooper-Schriefer) model of superconductivity
and the GL 
(Ginzburg-Landau) model of second-order transition.
The BCS model is the basic microscopic model of electrons and
its variables are electron operators $C_\sigma(x)$ at 
spatial point $x$,
while the GL model is the phenomenological model,
and its variables are an 

\vspace{0.5cm}
\begin{center}
\postscriptbox{8cm}{4cm}{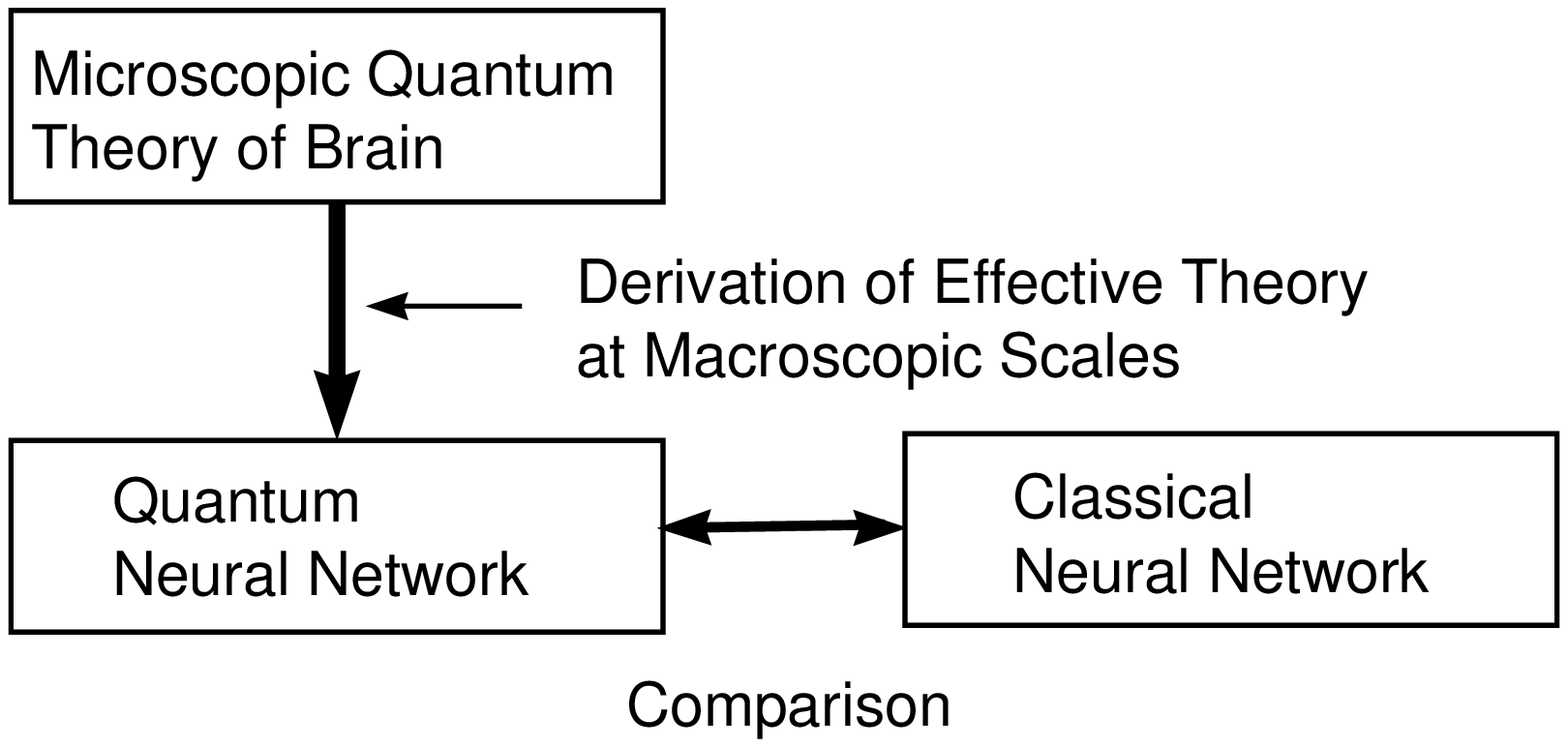}
\end{center}

\vspace{0cm}
Fig.2.
\ Relation between various models. To study the quantum
effects upon human brain, one should compare classical neural network models
and quantum neural network models. The latter is derived from the 
underlying microscopic quantum models as effective
(phenomenological) models at macroscopic scales. 

\label{fig:comp}

\noindent
order-parameter field $\phi(x)$, i.e.,
complex numbers that describe quantum amplitudes of Cooper pairs
of electrons. The relation between two sets of variables are
\begin{eqnarray}
\phi(x) & = & \langle C_{\uparrow}(x)C_{\downarrow}(x)\rangle, 
 \end{eqnarray}
where the brackets implies a statistical average over
the canonical ensamble at temperature $T$.

The GL theory was originally introduced as a phenomenological
model of superconductivity, but now one can derive it from 
the BCS model systematically as its effective model by using path-integral 
techniques\cite{sakita}. In fact, one starts from the BCS 
Hamiltonian $H_{\rm BCS}$ to obtain the GL free energy 
$F_{\rm GL}$ as
\begin{eqnarray}
Z_{\rm BCS} & \equiv& {\rm Tr}\ e^{-\beta H_{\rm BCS}(C)}\nonumber\\
&=& \int [dC] e^{A_{\rm BCS}(C)} 
 = \int [dC][d\phi] e^{A(C,\phi)}\nonumber\\
 &\simeq& \int [d\phi] e^{-\beta F_{\rm GL}(\phi)},
\end{eqnarray}
where $\beta = 1/(k_{\rm B} T)$ and the complex field $\phi(x)$ 
is introduced 
as an auxiliary field (integration variable) via the Hubbard-
Stratonovich transformation.\cite{sakita}

Because the GL theory may be viewed as a prototype of our
neural-netrwork model of Sect.\ref{u1gaugemodel}, 
let us explain it in some detail.
If we consider the system in a magnetic field,
$F_{\rm GL}$ is written  (we set $\hbar = c = 1$) as
\begin{eqnarray}
F_{\rm GL} = \int d^3x\left[
|D_{\mu}\phi|^2 + a |\phi|^2 + b |\phi|^4 +
\frac{1}{4} F_{\mu\nu}F_{\mu\nu}
\right]
\label{glfreeenergy}
\end{eqnarray}
where  $\mu = 1,2,3$ is the direction index, 
$F_{\mu\nu} = \partial_{\mu}A_{\nu} - \partial_{\nu}A_{\mu}$
is the magnetic field, $A_\mu(x)$ is the vector potential
(connection of gauge theory), and
$D_{\mu} = \partial_{\mu} -2ieA_{\mu}$ is the covariant derivative.
The coefficients $a = \alpha (T - T_c),\ \alpha > 0$, $b$, and
the critical temperature $T_c$ are calculable
and expressed by the parameters of the BCS model.
$F_{\rm GL}$ is invariant under the  
local  $U(1)$ gauge transformation,
\begin{eqnarray}
\phi(x)&\rightarrow& \phi'(x) = e^{2ie \alpha(x)} \phi(x),
\nonumber\\
A_{\mu}(x)&\rightarrow& A_{\mu}'(x) = A_{\mu}(x) 
+\partial_{\mu}\alpha(x).
\end{eqnarray}
The order parameter in zero magnetic field behaves as 
\begin{eqnarray}
\langle \phi(x) \rangle &= & \left\{
\begin{array} {cc}
\displaystyle \left[\frac{\alpha(T_c-T)}{2b}\right]^{1/2}, & T \leq T_c,\\
0, & T_c < T.
\end{array}
\right.
\end{eqnarray}
The equations of motion are obtained from 
\noindent
$\delta F_{\rm GL}/\delta \phi(x)$
$ = 0,$ 
$\delta F_{\rm GL}/\delta A_{\mu}(x) = 0$ as
\begin{eqnarray}
&& \big(-D_\mu D_\mu +a + 2 b |\phi|^2 \big)\phi = 0,\nonumber\\
&& \vec{\nabla} \times  \vec{B} = \vec{j}, \ 
\vec{B} =  \vec{\nabla} \times \vec{A}, \nonumber\\
&& \vec{j} = -2i e ( \bar{\phi}\ \vec{\nabla} \phi 
- \vec{\nabla} \bar{\phi}\  \phi ) - 8e^2 |\phi|^2 \vec{A}.
\label{equationofmotion}
\end{eqnarray}

Thus, we seek for an effective neural-network model 
that corresponds to the GL model of the BCS model, although
we don't specify the details of the underlying 
quantum model of the brain. 
Explicitly, we introduce a  $U(1)$ variable 
$S_i = \exp(i\varphi_i) \in U(1)$ to describe
the  quantum state of the $i$-th neuron and a 
$U(1)$ variable $J_{ij} = \exp(i\theta_{ij}) \in U(1)$
to describe the quantum state of the axon 
(synaptic connection) connecting $i$ and $j$-th neurons.
Physically, $S_i$ may be viewed as a wave function
of the $i$-th neuron which is in a quantum-mechanical coherent state 
 of its microscopic constituents, i.e., electrons, chemical materials,
and so on.
Likewise, $J_{ij}$ is a wave function of a coherent
state of the axon. We note that the idea that an axon
is well expressed by a single coherent state
is consistent with the theory of Hameroff and 
Penrose.\cite{hameroffandpenrose}

Furthermore, the requirement of gauge symmetry
is naturally incorporated into this assignement of $U(1)$
variables by regarding $S_i$ as a charged matter field
 and $J_{ij}$ as an exponentiated gauge connection.

\subsection{U(1) Gauge Model}
\label{u1gaugemodel}

Let us formulate the model on a three-dimensional cubic lattice.
We specify each site by the site-index $x$ and use $\mu = 1,2,3$
as the direction index. We use $\mu$ also as the unit vector in the
$\mu$-th direction. We set the lattice spacing $a = 1$ for simplicity. 
As explained in Sect.2.2,  for each site $x$ we put a $U(1)$ variable, 
\begin{eqnarray}
S_x = \exp(i\varphi_x),\ 
\end{eqnarray}
and for each link $(x\mu) \equiv (x,x + \mu)$, i.e., 
for nearest-neighbor (NN) pair of sites,
we  put another $U(1)$ variable,
\begin{eqnarray}
J_{x\mu} = \exp(i\theta_{x\mu}).
\end{eqnarray}
The local gauge transformation is expressed as 
\begin{eqnarray}
S_{x} &\rightarrow& V_x S_x,\nonumber\\
J_{x\mu} &\rightarrow&  V_{x+\mu}U_{x\mu}\bar{V}_{x},
\label{gaugetrsf}
\end{eqnarray}
where $V_x = \exp(i\alpha_x)$ and the bar implies the complex 
conjugate (e.g., $\bar{S}_{x} \equiv \exp(-i\varphi_x)$).

We propose the following gauge-invariant energy $E$;
\begin{eqnarray}
E &=& -\frac{c_1}{2} \sum_x \sum_{\mu}\big( \bar{S}_{x+\mu} 
J_{x\mu} S_x + {\rm c.c} \big)\nonumber\\
&& -\frac{c_2}{2} \sum_{x} \sum_{\mu > \nu} \big(
\bar{J}_{x\nu} \bar{J}_{x+\nu,\mu} J_{x+\mu ,\nu}J_{x\mu} 
+ {\rm c.c} \big)\nonumber \\
&& -\frac{c_3}{2} \sum_x \sum_{\mu} \sum_{\nu(\neq \mu)}
\big( \bar{S}_{x+\mu}\bar{J}_{x+\mu,\nu} J_{x\nu} J_{x+\nu,\mu} 
S_{x} \nonumber\\
&&  
+\bar{S}_{x+\mu} J_{x-\nu+\mu,\mu} J_{x-\nu,\mu} 
\bar{J}_{x-\nu,\nu} S_{x} + {\rm c.c} \big).
\label{energy}
\end{eqnarray}

\begin{center}
\postscriptbox{6cm}{6cm}{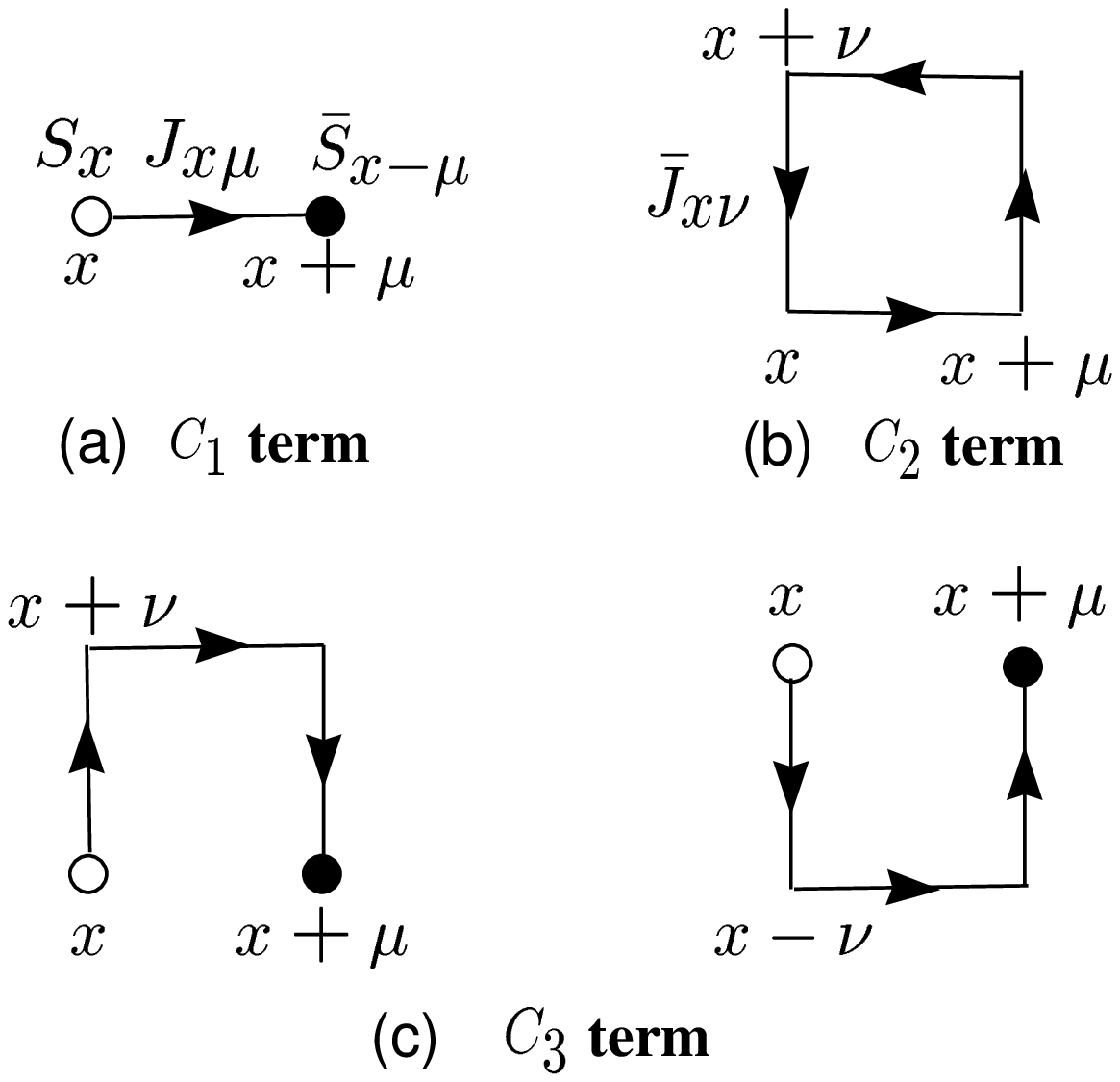}
\end{center}
\vspace{-0.5cm}
Fig.3. Graphical representation of each term in the energy $E$ 
of (\ref{energy}). Open circles for $S_x$.
Filled circles for $\bar{S}_x$. Straight lines  for $J_{x\mu}$.
The arrows distinguish $J_{x\mu}$ and $\bar{J}_{x\mu}$. The gauge 
invariance requires the lines  with arrows 
should (i) start from open circles,
(ii) end at filled circles, (iii) continue in a single direction.
\vspace{0.5cm}

\noindent
Each term in (\ref{energy}) is depicted in Fig.3.
$E$ of (\ref{energy}) reduces to the energy
of the $Z(2)$ model\cite{kemuriyamaandmatsui} if we replace 
$S_x, J_{x\mu}$ by $Z(2)$ variables.

Let us discuss the continuum limit ($a \rightarrow 0$) of $E$. 
Following Wilson\cite{wilson}
we write $\theta_{x\mu}$, the phase of $J_{x\mu}$, as
$\theta_{x\mu} = g a A_{\mu}(x)$  where $g$ is the gauge coupling 
constant, $a$ is t. To take the limit $a \rightarrow 0$,
we expand $J_{x\mu}$ 
w.r.t. $a$ as $J_{x\mu} = \exp(igaA_{\mu}(x)) \simeq 1 + igaA_{\mu}(x)
-g^2 a^2 A_{\mu}(x)^2/2 +O(a^3)$. Also we scale 
$S_{x} \propto a^{1/2}\phi(x)$. Then, by taking $c_i$ appropriately,
we find 
\begin{eqnarray}
c_1, \ c_3\ {\rm terms} &\rightarrow& |D_\mu \phi|^2\ {\rm and}\ |\phi|^2\
 {\rm terms}, 
\nonumber\\
 c_2\   {\rm term} &\rightarrow& F_{\mu\nu}F_{\mu\nu}\ {\rm term}.
\label{econtinuum}
\end{eqnarray} 
Although we introduced the $U(1)$ GL theory (\ref{glfreeenergy}) 
just as a typical example of an effective theory of a microscopic 
quantum model, it now serves as a continuum limit of the present 
lattice model.\cite{continuum}  
One can draw some useful informations using this relation.
For example, one may obtain a 
``current" $j_{x\mu}$ on the lattice as
\begin{eqnarray}
j_{x\mu} &=&\frac{\delta E|_{c_1 {\rm and } c_3 {\rm terms}}}
{\delta \theta_{x\mu}}\nonumber\\
&=&
 -\frac{i}{2}\Big(c_1 \bar{S}_{x+\mu} 
J_{x\mu} S_x  - {\rm c.c.}\Big) +\cdots,
\label{current}
\end{eqnarray}
which reduces to $\vec{j}(x)$  of (\ref{equationofmotion})
in the continuum limit. $j_{x\mu}$ is gauge invariant and may be
useful to describe the state of the system.

\subsection{Time Evolution}
\label{ssec:timeevolution}

Let us consider the dynamics of $S_x(t)$ and $J_{x\mu}(t)$.
As in the Hopfield model and Z(2) gauge model, we let
the energy $E$ basically decreases as 
the time increases with some rate of failures.  
These failures are caused by misfunctioning of signal processings
due to noises, etc.,
and may be controlled by the ``temperature" $T$; For higher(lower) $T$, 
failures occur more(less). This $T$ should not be confused
with the physical temperature of the brain, although there may be
some correlations among them.  


As explicit rules of time evolution, the following two are
possible;\\

\noindent
{\bf (I) Metropolis algorithm  (MA):}

MA\cite{metropolis} is a standard algorithm 
to calculate the thermal averages 
$\langle O(\{S_x\}, \{J_{x\mu}\}) \rangle$
over Boltzmann distribution,
\begin{eqnarray}
\langle O \rangle &=& 
\frac{1}{Z} \int[dS][dJ]\ O\ 
\exp(-\beta E),
\label{thermalaverage}
\end{eqnarray}
by generating a Markov(stochastic) process 
$\{S_x(\ell\Delta t)\}$, 
$\{J_{x\mu}(\ell\Delta t)\}\ 
(\ell= 1,2,\cdots,M)$ as 
\begin{eqnarray}
\hspace{-1cm}
\langle O \rangle  &=& \lim_{M \rightarrow \infty }
\frac{1}{M} \sum_{\ell=1}^{M} O(\{S_x(\ell\Delta t)\}, 
\{J_{x\mu}(\ell\Delta t)\}).
\label{markov}
\end{eqnarray}
By identifying $\ell\Delta t$ as the real time $t$, 
one may use this Markov process 
$\{S_x(\ell\Delta t)\}, \{J_{x\mu}(\ell\Delta t)\}$ itself 
just as their  time evolutions as proposed in the $Z(2)$ gauge 
model.\cite{matsui, kemuriyamaandmatsui}
The rates of changes in variables are controllable
by adjusting some parameters contained in MA.\cite{kemuriyamaandmatsui}
In particular, $S_x$ and $J_{x\mu}$ may have different rates.
If $J_{x\mu}$ change much slower than $S_x$, it may be more suitable
to first take an ensamble average over $S_x$ for fixed $J_{x\mu}$
and then take average over different $J_{x\mu}$ as in the 
theory of spin glass.\cite{spinglass}\\
 
\noindent 
{\bf (II) Langevin equation:}

Langevin equations\cite{langevin} are stochastic equations for
continuous variables. 
Since $U(1)$  variables are constrained (e.g., $\bar{S}_x S_x = 1$), 
it is preferable to focus on their phases, i.e. angles (mod$(2\pi)$) 
as independent variables. Then  one has
\begin{eqnarray}
\alpha_{\varphi}\frac{d{\varphi}_x}{dt} &=& 
-\frac{\partial E}{\partial \varphi_x} + \sqrt{2T} \eta_{\varphi_x},
\nonumber\\
\alpha_\theta\frac{d{\theta}_{x\mu}}{dt} &=& 
-\frac{\partial E}{\partial \theta_{x\mu}} + \sqrt{2T} 
\eta_{\theta_{x\mu}},
\label{langevin}
\end{eqnarray}
where $\alpha_{\varphi, \theta}$ are parameters to fix the time scales, 
and $\eta_{\varphi,\theta}$ 
are random white noises specified by their averages, 
\begin{eqnarray} 
\langle \eta_a(t) \rangle &=& 0,\nonumber\\
\langle \eta_a(t_1) \eta_b(t_2)\rangle  &=& \delta_{ab} \delta(t_1 - t_2).
\end{eqnarray}

In the energy $E$,
the term $c_1$, which corresponds to the energy of the Hopfield model,
describes the direct transfer of signal from $x$ to $x+\mu$.
The term $c_2$ describes the self energy after the transfer of signal
through the contour $(x \rightarrow  x+\mu \rightarrow x+\mu +\nu 
\rightarrow x+\nu \rightarrow x)$.
It may express the  energy of circular currents.
The term $c_3$ describes indirect transfers of signal from $x$ to $x+\mu$
via the bypath, $(x \rightarrow  x+\nu \rightarrow x+\nu +\mu \rightarrow 
x$). 

At first, it may look strange that there appear 
the $c_2$ and $c_3$ terms in $E$, which contain direct contacts (products) 
of two connection variables like $J_{x\mu}$ and $J_{x+\mu,\nu}$,
because each synapse connection necessarily contacts with a neuron
but not with a neaby synapse. However, two successive transfers like
$S_x \rightarrow S_{x+\mu}$ and $S_{x+\mu} \rightarrow S_{x+\mu+\nu}$
are described as a product of corresponding factors as
\begin{eqnarray}
&&\bar{S}_{x+\mu+\nu}J_{x+\mu,\nu}S_{x+\mu} 
\times \bar{S}_{x+\mu}J_{x\mu}S_x \nonumber\\
&=& \bar{S}_{x+\mu+\nu}J_{x+\mu,\nu}J_{x\mu}S_x
\end{eqnarray}
due to $S_{x+\mu}\bar{S}_{x+\mu} =1$. 
This explains why the terms like $c_2$ and $c_3$-terms
may appear in $E$. Another explanation 
is given in Ref.\cite{kemuriyamaandmatsui} based on the 
renormalization group.


\section{Phase Structure}
\label{sec:phase}

\subsection{Mean Field Theory}
\label{ssec:mft}

The MFT may be formulated as a variational method\cite{feynman}
fot the Helmholtz free energy $F$;
\begin{eqnarray}
Z& =& \int [dS] [dJ]
\exp(-\beta E) \equiv \exp(-\beta F),\nonumber\\
\hspace{-1.5cm}
\int [dS] &\equiv&  \prod_x \int_{0}^{2\pi}\frac{d\varphi_x}{2\pi},\ 
\int [dJ] \equiv \prod_{x\mu}\int_{0}^{2\pi} 
\frac{d\theta_{x\mu}}{2\pi}. 
\label{partitionfunction}
\end{eqnarray}
For a variational energy $E_0$ there holds the following relations;
\begin{eqnarray}
Z_0& =& \int [dS]  [dJ] 
\exp(-\beta E_0) \equiv \exp(-\beta F_0),\nonumber\\
F &\le& F_v \equiv F_0 + \langle E -E_0 \rangle_0, \nonumber\\ 
\langle O \rangle_0 &\equiv& Z_0^{-1} 
\int [dS]  [dJ] \ O\ 
\exp(-\beta E_0).
\end{eqnarray}
From this Jensen-Peierls inequality, we adjust the variational parameters
contained in $E_0$ so that $F_v$ is minimized.

For the trial energy $E_0$ of the present system,
we assume the translational invariance
and consider the following sum of single-site and single-link energies;
\begin{eqnarray}
E_0 = - W \sum_{x\mu}J_{x\mu} - h \sum_x S_x,
\end{eqnarray}
where $W$ and $h$ are real variational parameters.
Then we obtain the following free energy per site, $f_v \equiv F_v/N$,
where $N$ is the total number of lattice sites 
(We present the formulae for $d$-dimensional lattice);  
\begin{eqnarray}
f_v &=& -\frac{d}{\beta} \ln I_0(\beta W) -\frac{1}{\beta} 
\ln(I_0(\beta h) -c_1 dm^2 p  \nonumber \\
& -&c_2 \frac{d(d-1)}{2} p^4 - 2c_3 d(d-1) 
m^2 p^3 + dWp + hm,\nonumber\\
m &\equiv& \langle S_x 	\rangle_0 =  
\frac{I_1(\beta h)}{I_0(\beta h)},\ \  
p \equiv \langle J_{x\mu} 	\rangle_0 = 
\frac{I_1(\beta W)}{I_0(\beta W)},
\label{fv}
\end{eqnarray}
where $I_n(\gamma)\ (n:$integer) is the modefied Bessel function,
\begin{eqnarray}
I_{n}(\gamma) &=& \int_{0}^{2\pi} \frac{d\theta}{2\pi}
\exp(\gamma \cos\theta + i n \theta), 
\end{eqnarray}

The stationary conditions for $f_v$ w.r.t. $W, h$ read
\begin{eqnarray}
W &=& c_1 m^2 +2c_2 (d-1)p^3 +6 c_3 (d-1)m^2 p^2, \nonumber\\
h &=& 2dc_1 mp +4c_3 d(d-1)mp^3.
\label{mft}
\end{eqnarray}
For many systems, the MFT is known to become exact for $d \rightarrow 
\infty$. It is proved also for the $Z(2)$ model ($c_3 = 0$) \cite{drouffe}
by assuming suitable scaling behaviors of parameters $\beta c_i$
at large $d$.

\subsection{Phase Structure}
\label{ssec:phase}

The MFT equations (\ref{fv}-\ref{mft}) for $d=3$ generate 
the three phases characterized in the following Table1;\\
\begin{center}
\begin{tabular}{|c|c|c|c|} 
\hline
   phase    & $\langle J_{x\mu} \rangle $ & $\langle S_x \rangle $  
   &  ability 
\\ \hline
Higgs       & $\neq 0$  & $\neq 0$  & learn and recall 
\\ \hline
Coulomb     & $ \neq 0$  & $0$  & learn 
\\ \hline
Confinement & $0$   & $0$     &  N.A.
\\
\hline
\end{tabular}

\vspace{0.3cm}
Table1. Phases and order parameters.
\end{center}
\vspace{0.5cm}

\noindent
In the first column of Table1,  
the name of each phase is given, which
are used in particle physics. The second (third) column 
shows the order parameter $\langle J_{x\mu}  
\rangle = p$  ($\langle S_{x} \rangle = m$).
The fourth column shows the properties of each phase
characterized by these order parameters.\cite{matsui,kemuriyamaandmatsui}
The condition $p \neq 0$ is a necessary condition
to learn a pattern of $S_x$ by storing it to $J_{x\mu}$,
while $m \neq 0$ is  a necessary condition to recall it
as in the Hopfield model.
We note that the combination $p = 0$ and $m \neq 0$ is missing.

In Fig.4, we plot the phase boundaries 
obtained from
(\ref{fv}-\ref{mft}) by solid curves for various values of $c_3$.
We have also superposed the MFT results of the $Z(2)$ gauge model
\cite{matsui, kemuriyamaandmatsui} by dashed curves for comparison.

\newpage

\begin{center}
\postscriptbox{7cm}{4.5cm}{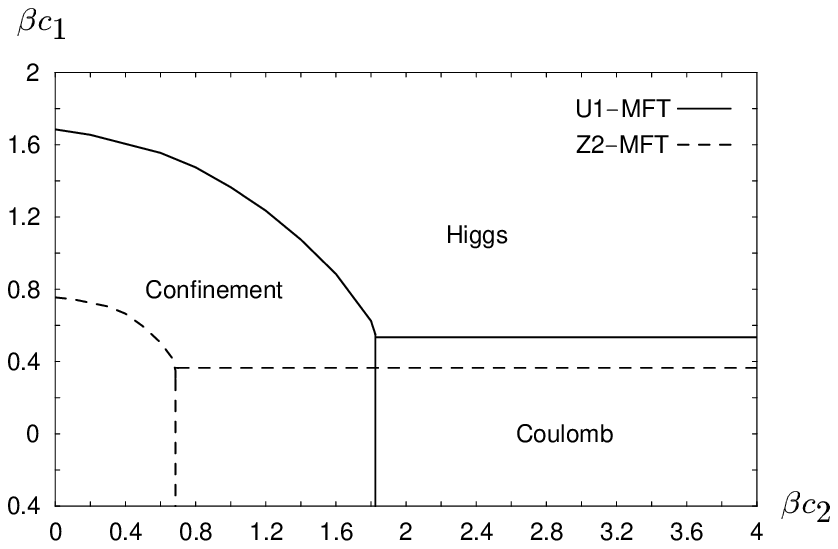}

(a) \ $\beta c_3 = -0.05$
\end{center}

\begin{center}
\postscriptbox{7cm}{4.5cm}{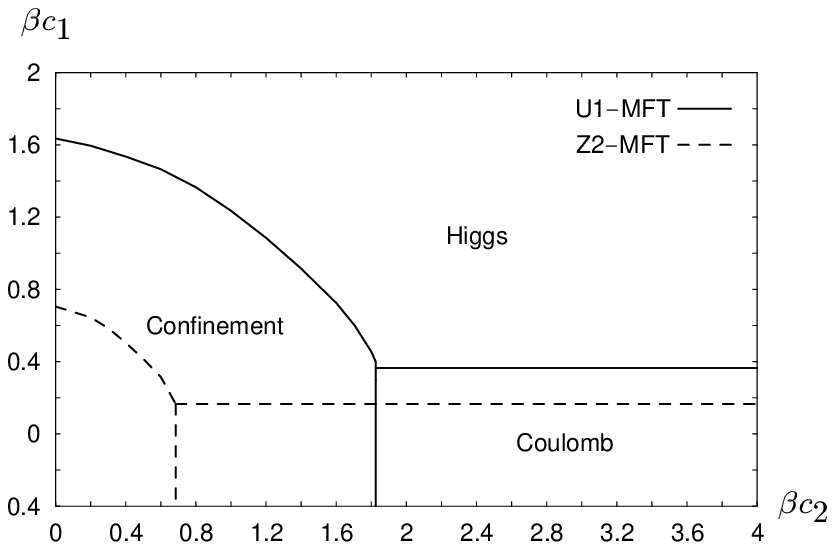}

(b) \ $\beta c_3 = 0.0$
\end{center}

\begin{center}
\postscriptbox{7cm}{4.5cm}{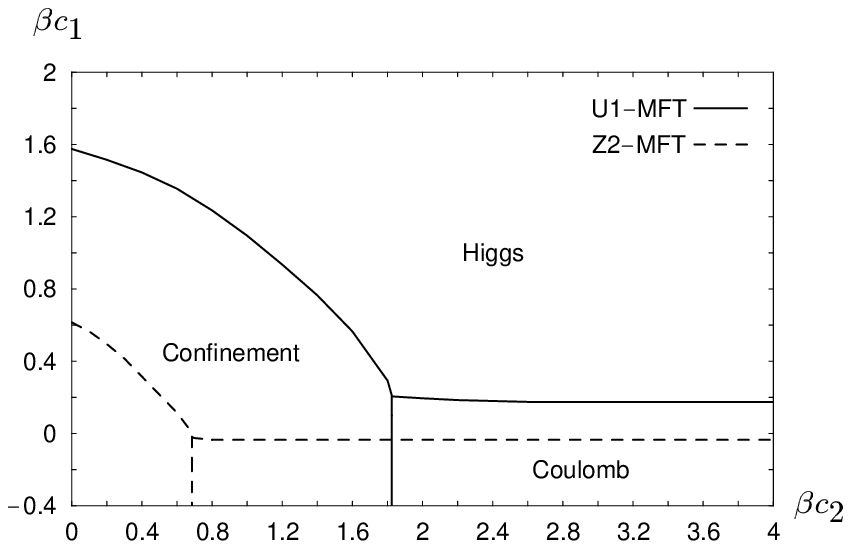}

(c) \ $\beta c_3 = 0.05$
\end{center}

\begin{center}
\postscriptbox{7cm}{4.5cm}{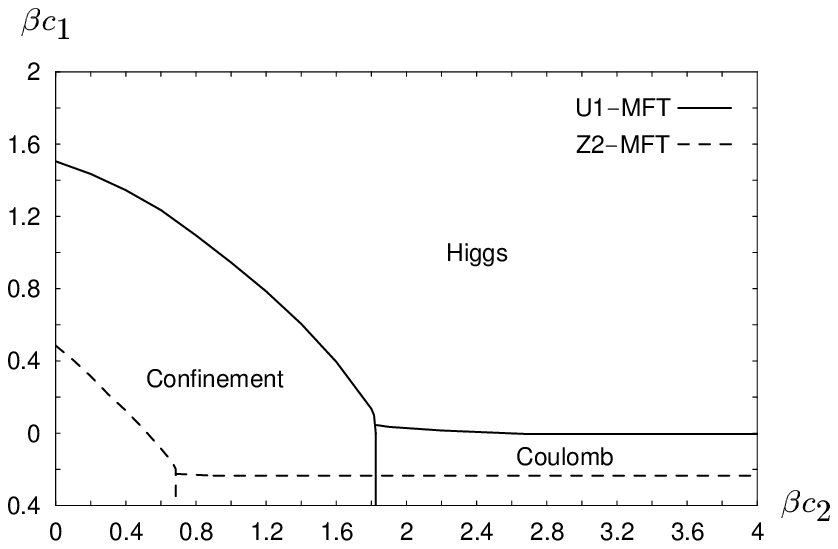}

(d) \ $\beta c_3 = 0.1$
\end{center}
Fig.4. Phase diagrams in MFT. Dashed lines represent
the phase boundaries in  MFT of $Z(2)$ 
model.\cite{kemuriyamaandmatsui}




The phase boundary of MFT between Higgs phase and Coulomb
phase is second order, while other two boundaries, Higgs-confinement 
and confinement-Coulomb, are first order.
Across a second-order transition, $p$ and $m$ vary continuously,
while across a first-order transition, $p$ and/or $m$ change 
discontinuously with finite jumps of $\Delta p$ and/or
$\Delta m$. For a Higgs-confinement transition,
$\Delta p \neq 0$ and $\Delta m \neq 0$, and
 for a confinement-Coulomb transition,
 $\Delta p \neq 0$ and  $\Delta m = 0$ since $m = 0$ in both phases.

The present MFT may have some inappropriate points, which should 
be improved by more accurate methods like Monte Carlo 
simulations;

(i)  In the pure gauge case where $c_1 = c_3 = 0$,
the system is known to support always the confinement 
phase for any values of 
$c_2$.\cite{polyakov} This indicates that 
the confinement-Coulomb transitions
may disappear in some parameter regions (e.g., for $c_3 = 0$). 
Then the confinement phase should survive and the Coulomb phase
 should disappear.

(ii) Along the correct Higgs-confinement boundary,
the jumps $\Delta p, \Delta m$ may decrease as $c_2$ decreases 
and disappear at a certain point with $c_2 > 0$. 
This critical point correponds to the complementarity studied in 
Ref.\cite{complementarity} for $c_3 = 0$, which states that
 these two phases are analytically connected via a detour.

Let us next comment on the Elitzur's theorem.\cite{elitzur}
It states that expectation values of {\it gauge-variant} 
objects should vanish. 
Thus $\langle S_x \rangle = \langle J_{x\mu} \rangle =0$.
This sounds to prohibit deconfinement phases like Higgs phase and Coulomb
phase in Table 1. 
However, these deconfinement phases certainly exist.
To compromise the MFT results with the
Elitzur' theorem, one just needs to  average over gauge-rotated
copies of a MF solution\cite{drouffe}.
Actually, the solution $m = \langle S_x \rangle, 
p = \langle J_{x\mu} \rangle$ is degenerate in the free energy 
with their gauge copies
 $\langle S'_x \rangle$ and $\langle J'_{x\mu} \rangle$, 
 and should be superposed to satisfy the Elitzur's 
 theorem. The location and the nature of phase transitions are unchanged.

\section{Discussion}
\label{sec:discussion}

We have proposed a quantum model of neural network based on 
gauge principle. The model resembles lattice $U(1)$ Higgs gauge theory,
exhibiting a rich phase structure. The model should be regarded
as a phenomenological (effective)  model of an underlying 
microscopic quantum theory of the brain in the sense that 
the variables of the model, $S_x$ and $J_{x\mu}$,
describe coherent quantum states of each neuron at $x$ and 
axon (or synaptic connection) along $(x,x+\mu)$, respectively.

We have not specified the underlying microscopic theory,
although there are some candidates.\cite{umezawaandtakahashi,
hameroffandpenrose, matsuiandsakakibara}
This point {\it is not} a flaw but an advantage
since the essential characteristics
of the effective model at low energies are to be determined by only
a few properties of the microscopic model like
dimensionality, symmetry, etc. This is known 
as the universality in renormalization group. 
The present $U(1)$ model will apply for a wide variety
of microscopic models describing ``charged" particles
and gauge bosons in three dimensions with local $U(1)$ gauge
symmetry. The model  of Stuart, Takahashi and 
Umezawa\cite{umezawaandtakahashi} is such a model.
Also the model of a microtubule proposed 
in Ref.\cite{matsuiandsakakibara}
can be cast into this category because the Coulomb
interactions among electrons can be written as gauge
interactions mediated by gauge bosons, i.e., photons.

The model may be regarded as an extension of the classical $Z(2)$
gauge model\cite{matsui,kemuriyamaandmatsui} to the gauge group $U(1)$. 
This similarity makes it easy to compare these two models
and single out the difference between them, which is to be 
interpreted just as the quantum effects. 

On the level of phase structure in MFT,
Fig.4 shows that the region of the confinement phase in the 
$U(1)$ model is wider than that of $Z(2)$ model. 
This is due to quantum fluctuations;
the  $U(1)$ variables are continuous while $Z(2)$ variables are
discrete. In short, the critical temperatures
(both $c_1$ and $c_3$) of the $U(1)$ model
is higher than those of the $Z(2)$ model. 
From the Table 1, 
this implies that the ability of learning patterns and
recalling them is weakened globally by the quantum effects.
More detailed study of this point is to be done in simulations of
individual learning and recalling processes by using 
the rule of time evolution in Sect.\ref{ssec:timeevolution}.

Another significant difference is that $U(1)$ model
allows us to define the current $j_{x\mu}$ of (\ref{current}) 
as in $\vec{j}(x)$ of (\ref{equationofmotion}). This is possible
because the $U(1)$ gauge symmetry is not discrete but continuous.
For a system with a continuous symmetry, one may obtain conserved current
by applying Noether's theorem. 
It is worth to mention the difference between
the present $U(1)$ gauge variables $\theta_{x\mu}$,
the exponent of $J_{x\mu}$, and
the vector potential $A_\mu(x)$ in (\ref{glfreeenergy}). 
Although both are gauge fields,
$A_\mu(x)$ describes the usual electromagnetic field,
while $\theta_{x\mu}$ describes the synaptic connections.
They are independent each other. Thus, $j_{x\mu}$ is {\it not}
the electromagnetic current.  We need to scrutinize
the  physical meaning of $j_{x\mu}$ further,  although one
expects that it describes the flows of electric voltage along axons 
and accompanying chemical materials at synaptic connections.

Let us comment here on the usefulness of such current for 
another network models. In some models that have real continuous
$J_{ij} (\in (-\infty, \infty))$, $J_{ij}$ diverges to $\pm \infty$ 
as time runs. 
Without imposing artificial and unnatural conditions 
to avoid divergences of $J_{ij}$, a conserved current, i.e., local continuous
 gauge symmetry, may assure us that $J_{ij}$ shall not diverge,
 since the total amount of chemical materials are finite.

In the present lattice model,
the gauge-invariant current $j_{x\mu}$ can be used to scan
the network at every time step. By monitoring
 $j_{x\mu}$ during the processes of learning
and recalling, one may  study the activities
of network as quantum transports systematically.
This is an interesting subject in future.

Finally, let us list up other possible problems in future study.

\noindent
- More realistic phase strucutre by Monte Carlo simulaitons.

\noindent
- Simulation of  processes of learning
patterns and recalling them through the time evolution
in Sect.\ref{ssec:timeevolution}.

\noindent
- Inclusion of long-range interactions into the energy.

\noindent
- Introduction of another set of gauge variables $\tilde{J}_{x\mu} $
to study the effect associated with the asymmetric couplings
$J_{ij}$ and $J_{ji} (\neq J_{ij})$ \cite{matsui}, 
which is reflected  by $\tilde{J}_{x\mu} \neq \bar{J}_{x\mu}$.\\ 

{\bf Acknowlegement}\\

We thank Professor Kazuhiko Sakakibara for interesting discussions
and suggestions.\\



\begin{thebibliography}{99}
\bibitem{gauge}
Strictly speaking, the connection $A_{xy}$ connects two nearby points
$x,y$. $U_{xy}$ is the path-dependent phase factor
$\prod_{{\rm path}}\exp(iA_{ab})$ where the product is taken along
a specified path from $y$ to $x$.
\bibitem{continuum}
To obtain the $|\phi_x|^4$ term of $F_{\rm GL}$, one needs to
relax the condition $|S_x| =1$ and include $|S_x|^4$ term in $E$. 
Such radial fluctuations $|S_x|$ are massive and irrelevant
in low-energy physics.
\end{thebibliography}
\end{document}